\documentclass[aps]{revtex4}
\usepackage{graphicx}
\usepackage[all]{xy}
\usepackage{amsmath}
\usepackage{amssymb}
\usepackage{color}

\newcommand{\bb}{\bibitem}
\newcommand{\bes}{\begin{subequations}}
\newcommand{\ees}{\end{subequations}}
\newcommand{\benn}{\begin{eqnarray*}}
\newcommand{\eenn}{\end{eqnarray*}}

\newcommand{\be}{\begin{equation}}
\newcommand{\ee}{\end{equation}}
\newcommand{\ben}{\begin{eqnarray}}
\newcommand{\een}{\end{eqnarray}}

\def\ben{\begin{eqnarray}}
\def\een{\end{eqnarray}}
\def\be{\begin{equation}}
\def\ee{\end{equation}}

\begin{document}

\title{Four-dimensional gravity on supersymmetric dilatonic domain walls}
\author{ R.C. Fonseca$^{2,3}$, F.A. Brito$^{1}$ and L. Losano$^{1,3}$}
\affiliation{$^{1}$ Departamento de F\'\i sica,
Universidade Federal de Campina Grande, Caixa Postal 10071,
58109-970  Campina Grande, Para\'\i ba, Brazil\\
$^{2}$Departamento de F\'\i sica, Universidade Estadual da Para\'\i ba, 58.109-753 Campina Grande, PB, Brazil\\
$^{3}$ Departamento de F\'\i sica, Universidade Federal da Para\'\i ba, Caixa Postal 5008, 58051-970 Jo\~ao Pessoa, Para\'\i ba, Brazil}

\vspace{3cm}

\date{\today}

\begin{abstract}
We investigate the localization of four-dimensional metastable gravity in supersymmetric dilatonic domain walls through massive modes by considering several scenarios in the model. We compute corrections to the Newtonian potential for small and long distances compared with a crossover scale given in terms of the dilatonic coupling. 4D gravity behavior is developed on the brane for distance very much below the crossover scale, while for distance much larger, the 5D gravity is recovered. {Whereas in the former regime gravity is always attractive, in the latter regime due to non-normalizable unstable massive graviton modes present on the spectrum, in some special cases, gravity appears to be repulsive and signals a gravitational confining phase which is able to produce an inflationary phase of the Universe.}
\end{abstract}

\maketitle

\section{Introduction}

One of the main interesting results in higher dimensional physics in recent years, is the possibility of localization of gravity on braneworlds which was initiated in the seminal paper  by Randall-Sundrum (RS) \cite{RS1}. This attracted great attention because it was the first example where four-dimensional gravity could be localized even in the presence of extra dimensions with infinite size. {Before this approach, the well-known Kaluza-Klein mechanism was the only largely considered alternative to achieve four-dimensional gravity from higher-dimensional theories by compactifying the extra dimensions into very small radii. On the other hand, braneworlds is an alternative to compactification, because a 3-brane embedded into higher dimensional space is able to localize four-dimensional gravity, even if the extra dimensions have infinite size. This mechanism works because the 3-brane is embedded in a curved higher dimensional space.} More specifically, in the original set up \cite{RS1}, a 3-brane with tension sourced by a delta function is embedded in a warped five-dimensional bulk with negative cosmological constant, { that is, an $AdS_5$ space}. The {relationship} between the brane tension and the bulk cosmological constant plays the role of producing a zero mode solution to the linearized Einstein equations.  This solution given in terms of the metric developing the suitable { fall-off with respect to the extra dimension}  is responsible for localizing the four-dimensional gravity. The following massive modes of the full spectrum also enter in the computation of the Newtonian potential describing corrections to the four-dimensional gravity. They should be highly suppressed at large distance but should deviate the Newton's law from the four-dimensional behavior at very small distance, implying that the five-dimensional character of the full space shows up at very high energy. 

Since the RS paper,  many extensions to thick 3-brane have appeared in the literature \cite{SDW} as domain wall solutions in five or higher dimensional  theories with scalar fields coupled to gravity. Most of them are conceived in such a way that can be thought of as the bosonic sector of a supergravity theory or at least the { bosonic sector} of the sometimes referred to as fake supergravity \cite{FS}. One of the major problems in realizing the RS set up in five-dimensional supergravity is the difficulty to find normalizable zero modes. This issue was explored several times in the literature --- see \cite{cvetic2000} for a review. The main point is that the warp factor found does not enjoy the required fall-off along the extra dimension. One can rephrase this problem in terms of the volume of the five-dimensional spacetime. In the RS scenario, in spite of the fact { that} the extra dimension is infinite, { the compactification from five down to four dimensions is ensured because the volume of the $AdS_5$ space is finite - contrary to Minkowski space whose volume is infinite}. 

{ However, soon after the appearance of the RS scenario}, it was presented alternative scenarios by Gregory-Rubakov-Sibiryakov (GRS) \cite{RGS} and Dvali-Gabadadze-Porrati (DGP) \cite{DGP}  where the volume of the five-dimensional spacetime is not necessarily  finite to localize gravity. This opened the possibility of localizing four-dimensional gravity in braneworlds embedded in  asymptotically flat five-dimensional spaces. The price to pay is { that} instead of zero modes the localization is due to massive graviton modes and the localization is metastable. In spite { of} its metastability, the  localization of metastable gravity can survive long enough if one probes distances very much { smaller} than a crossover scale. In these alternative scenarios, the brane is still sourced by a delta function but  the bulk in both cases is asymptotically five-dimensional Minkowski  space. { As it was previously stated \cite{RGS,DGP,FBL1,FBL2}, the crossover scale $r_c$ can be related to the Hubble size as $r_c\sim H^{-1}$. We shall mostly consider the large crossover regime $r_c\to\infty$ which is consistent with the present Hubble size, though, we shall also briefly comment on $r_c\to0$ that means the regime of the early Universe. }

Furthermore, have also been considered in the literature thick brane realizations of metastable localization of gravity in the literature \cite{KR,MM,SCH,ITO,PO0,BBG,NP,FBL1,FBL2}. In most of these cases one has considered five-dimensional theories of scalar fields coupled to gravity in the realm of `fake supergravity' where several issues were discussed.  One should, though, also consider such realizations in supergravity domain wall solutions \cite{CS}.

In the present study, we shall consider the supersymmetric dilatonic solution that can be found from a higher dimensional supergravity theory that after specific compactifications, such as toroidal compactification,  turns to a simpler five-dimensional theory of a scalar field coupled to gravity \cite{DI00,Di2,DI5,PO3}  in the same fashion of those conceived in \cite{SDW}. We shall investigate the possibility of localizing four-dimensional metastable gravity in supersymmetric dilatonic domain walls through massive gravitational modes by considering several scenarios, { including those in which non-normalizable unstable massive graviton modes determine dominant repulsive gravity.

The paper is organized as follows. In Sec.~\ref{SecII} we make a brief review on the mechanism of gravity localization on braneworlds. In Sec.~\ref{SecIII} we present four scenarios by properly adjusting one of the free parameters of the theory. In all of these cases is recovered the four dimensional attractive gravity in the regime $r\ll r_c$. The first case gives the expected result since the localized gravity is attractive everywhere. However, due to non-normalizable unstable massive graviton modes present on the spectrum, the following three cases assume a new behavior at large distance, i.e., at $r\gg r_c$. In this regime they exhibit a repulsive gravity with exponentially increasing potential which signals a gravitational confining phase that is able to produce an inflationary phase of the Universe \cite{linde}. This effect has also been addressed in earlier similar studies on time-like extra-dimensions \cite{chaichian, Matsuda:2000nk}. Finally in Sec.~\ref{SecIV} we make our final comments.}

\section{Gravity on dilatonic domain walls}\label{SecII}

 Let us briefly introduce the mechanism behind gravity localization in braneworlds. We consider a general dilatonic $(D-2)-$brane solution in a spacetime in $D$ dimensions discussed  in Refs.~\cite{DI00,Di2,DI5,PO3}. The spacetime is assumed to be the direct product $M_{D-1}\,\times\,K$ of the $(D-1)-$dimensional spacetime $M_{D-1}$ and some noncompact internal space $K$. We shall focus our attention to obtain solutions describing braneworlds as the worldvolume of domain walls. 
The Einstein-frame action for the dilatonic domain wall is
\ben\label{dW4}
S=\frac{1}{2\,\kappa_D^2}\int{d^Dx\,\sqrt{-g}\left({{R}}-\frac{4}{D-2}\left(\partial\phi\right)^2+\Lambda\,{\rm {e}}^{-2\,a\,\phi}\right)},
\een 
where $\kappa_D$ is the $D-$dimensional gravitational constant, and the cosmological constant term is multiplied by the
dilaton factor which depends on an arbitrary dilaton coupling parameter $a$ in any $D$-dimensional spacetime. The dilatonic domain wall solution given in terms of a conformally flat form is 
\be\label{dW5}
\bar{G}_{MN}dx^{M}dx^{N}=C(z)[-dt^2+dx_1^2+...+dx^2_{D-2}+dz^2],
\ee
\ben\label{dW6a}
\phi=\frac{(D-2)a}{2(\Delta+2)}\ln\left(1+\frac{\Delta+2}{\Delta}{Q}\,{|z|}\right)\,,\quad\quad C(z)=\left(1+\frac{\Delta+2}{\Delta}{Q}\,{|z|}\right)^{{4}/{(D-2)(\Delta+2)}},
\een
where
\ben\label{dW6b}
\Delta=\frac{(D-2)a^2}{2}-\frac{2(D-1)}{D-2}, \quad\quad {\rm{and}}\quad\quad \Lambda=-\frac{2\,Q^2}{\Delta}.
\een
For future discussions it is useful to define $\Lambda$ in terms of the space curvature $1/L$. In the limit of sufficiently small curvature ($L\gg \ell_s$) being $\ell_s$ the string length, the supergravity approximation takes place.
In order to study the spectrum of the graviton in the dilatonic domain wall
background, we consider the metric describing the small fluctuation
$h_{\mu\nu}(x^{\rho}, z)$ of the $(D-1)-$dimensional Minkowski spacetime embedded into the conformally flat
$D-$dimensional spacetime
\be\label{dW14}
\bar{g}_{\bar{\mu}\bar{\nu}}dx^{\bar{\mu}}dx^{\bar{\nu}}=C(z)[(\eta_{\mu\nu}+h_{\mu\nu})dx^{\mu}dx^{\nu}+dz^2],
\ee where $|h_{\mu\nu}|\ll1$. 

By using the linearized Einstein equations in the transverse traceless gauge, i.e., $\partial^{\mu}h_{\mu\nu}=0$,  and the
metric fluctuation in the form $h_{\mu\nu}(x^{\rho},z)=\bar{h}^{(m)}_{\mu\nu}(x^{\rho})C^{-(D-2)/4}\psi_{m}(z)$,  where $\eta^{\mu\nu}\partial_{\mu}\partial_{\nu}\bar{h}^{(m)}_{\mu\nu}=m^2\bar{h}^{(m)}_{\mu\nu}$, we obtain the Schroedinger-like equation
\be\label{sch1}
-\frac{d^2}{d\,z^2}\,\psi_{m}(z)+U(z)\,{\psi}_{m}(z)=m^2\,\psi_{m}(z),
\ee 
with the potential 
\be\label{dW20}
U(z)=\frac{D-2}{16}\left[(D-6)\left(\frac{C'}{C}\right)^2+4\,\frac{C''}{C}\right]=-\frac{Q^2\,(\Delta+1)}{\Delta^2\left(1+\frac{\Delta+2}{\Delta}\,Q|z|\right)^2}+\frac{2\,Q}{\Delta}\,\delta(z),
\ee where the prime denotes derivatives with respect to $z$. 
In this paper, we shall discuss the properties of the KK modes for
different values of $\Delta$ taking $Q > 0$. We explore the scenarios generated by the Schroedinger-like potentials that arises from four choices for the range of $\Delta$ , and we study the asymptotic behavior of the  corrections for the Newtonian potential, which are derived from the wave functions that emerge from the corresponding Schroedinger-like equations. This means to investigate how easy four-dimensional massive gravity can be localized on the brane.

Notice that the structure of this potential allows to write the
Schroedinger-like equation in a quadratic form \cite{SDW}
\be\label{fat}
H\,\psi(z)=m^2\,\psi(z),\quad H={\cal {Q^{\dagger}}{Q}},\quad {\cal Q}=-\frac{d}{dz}+\frac{D-2}{2}\,A'(z),
\ee where $A(z)=\frac12\,\ln\,C(z)$. The zero mode obeying $H\,\psi_0=0$ is given by
\be\label{Ze1}
\psi_0(z)=N_0\,{\rm{exp}}\left[\frac{D-2}{2}\,A(z)\right]=N_0\,C(z)^{\frac{D-2}{4}}=N_0\,\left(1+\frac{\Delta+2}{\Delta}\,Q|z|\right)^{1/(\Delta+2)}.\ee It is easy to conclude from the equation (\ref{Ze1}) that for $Q>0$ there exists normalizable zero mode only in the range $\Delta\leq-2$, \cite{DI5,PO3}. In the following, we consider four  cases which do not fall into this range, and do not exhibit zero mode.
{This is because we aim to do a complementary study of gravity localization on supersymmetric dilatonic domain walls by extending to massive modes the previously analysis mainly based on zero modes \cite{DI00,Di2,DI5,PO3}. This produces localization of four-dimensional gravity for distances very much smaller than a crossover scale $r_c$.}

\section{Four-dimensional gravity on  dilatonic domain walls}\label{SecIII}

\begin{figure}[ht]
\includegraphics[{height=04cm,width=08cm,angle=00}]{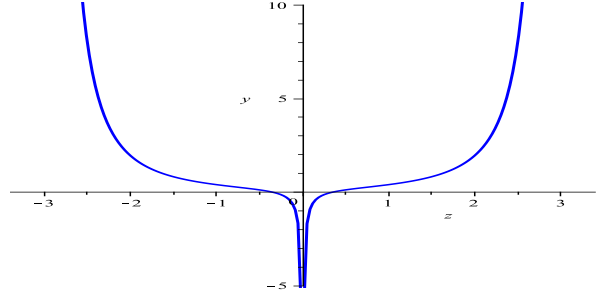}
\caption{The potential \eqref{BOX} as a function of $z$, for $\Delta=-3/2$, and $Q=1$.}\label{fig1}
\end{figure}

In the present section, we focus on four scenarios of 4D gravity on  dilatonic domain walls, by considering four ranges of $\Delta$, for $Q>0$.
\begin{itemize}
	\item i) $-2<\Delta<-1$ case.
\end{itemize}
In this case, the potential \eqref{dW20} can be written as
\be\label{BOX}
U(z)=\frac{{Q}^{2}}{{\Delta}^{2}}{\frac { \left|\Delta+1\right| }{\left(1-{\frac { \left(\Delta+2 \right) }{|\Delta|}}Q|z|\right)}}-\frac{2\,Q}{|\Delta|}\,\delta(z)\,,
\ee 
whose massive modes solution of the Schroedinger-like eq.\eqref{sch1} is given by \cite{STE}
\be\label{Bes12}
\psi_{m} \left( z \right) ={\rm{K}}_1(m)\,\sqrt {\frac{1}{\alpha_1}+ \left| z \right| }\,\left(J
_{{\beta_1}} \left( {\frac {m \left( 1+\alpha_1\, \left| z \right|  \right) 
}{\alpha_1}} \right) +F_1(m)\,N_{{
\beta_1}} \left( {\frac {m \left( 1+\alpha_1\, \left| z \right|  \right) }{
\alpha_1}} \right)\right) ,
\ee
in terms of the Bessel functions of first and second kind, $J_{\beta_1}$ and $N_{\beta_1}$ respectively, where $\alpha_1={\frac {(|\Delta|-2)\,Q}{|\Delta|}}$, $\beta_1=\frac12\,{\frac {\sqrt {8-8\,{ |\Delta|}+{{ \Delta}}^{2}}}{({|\Delta}|-2)}}$, and ${\rm{K}}_1$ and $F_1$ are  the normalization factors --- See Fig.~\ref{fig1}.\\
This problem is similar to that for a particle trapped into an infinite box with a delta function at the origin, which we use to impose the boundary conditions.
The jump condition at $z = 0$, is given by
\be\label{JUMP.fim2}
\left[{\frac {d}{dz}}{\psi}_{{m}} \left( z \right) \right
]_{{z=0}}=-\frac{2\,Q}{|\Delta|}\,{\psi}_{m}
 \left( 0 \right).
\ee
In order to obtain the correction term to the four-dimensional Newton's law
between two unit masses on the 3-brane ($D=5$), one needs to obtain the probability density of massive modes on the brane.
The asymptotic behaviors of $|{\psi}_{m}\left(0\right)|^2$ are obtained from the
normalization factors given by
\ben\label{NorFim2}
|{\rm{K}}_1{(m)}|^{2}=\frac{m}{2\,|\alpha_1|\,\left(1+F_4^{2}(m)\right)\,},\quad F_1 \left( m \right) = \frac{\left( -2\,J_{{\beta_1+1}} \left( {\frac {m}{  \alpha_1}} \right) m+\epsilon_{{1}}J_{{\beta_1}} \left( {\frac {m}{\alpha_1}  } \right)  \right) }{ \left( -2\,N_{{\beta_1+1}} \left( {\frac {m}{\alpha_1}  } \right) m+\epsilon_{{1}}N_{{\beta_1}} \left( {\frac {m}{\alpha_1}}   \right)  \right) },
\een
where $\epsilon_{{1}}=\frac{Q}{|\Delta|} \left( -|\Delta|+2\,\beta_{{1}} \left( |\Delta|-2 \right)\right)
$.
Here, we use the Lommel's formula, $J_{{\rm{n}}+1}(x)\,N_{{\rm{n}}}(x)-N_{{\rm{n}}+1}(x)\,J_{{\rm{n}}}(x)=2/\pi\,x$, to find the asymptotic forms of Bessel function as $J_{{{\rm{n}}}} \left( x \right) \sim\sqrt {{\frac {2}{x\pi }}}\cos
 \left( x-1/2\,{\rm{n}}\,\pi +1/4\,\pi  \right)
$ and $N_{{{\rm{n}}}} \left( x \right) \sim\sqrt {{\frac {2}{x\pi }}}\sin
 \left( x-1/2\,{\rm{n}}\,\pi +1/4\,\pi  \right)$ for $x\gg1$, $J_{{{\rm{n}}}} \left( x \right) \sim \left( 1/2\,x \right) ^{{\rm{n}}}
$ and $ N_{{{\rm{n}}}} \left( x \right) \sim-{ {1/\pi } \left( 1/2\,x \right) ^{-{\rm{n}}}}
$ for $x\ll1$.
Thus, the asymptotic probability density assumes the simplified forms
\ben\label{Prob1.fim2}
\left| {\psi}_{m} \left( 0 \right)  \right|^2&\sim&{\frac {\pi }{4\,{\alpha^{2}_1}}}\,\frac{{m}^{2}}{ \left( \epsilon_1^2+{ {4\,{m}^{2}}} \right)}
 ,\quad\,\,\,\quad\,\quad \quad\quad\quad\, \quad\quad\quad \quad\quad\quad \quad\quad\,m\gg\frac{1}{r_c};\\
 \left| {\psi}_{m} \left( 0 \right)  \right| ^{2}&\sim&  \left({\frac {{2\,|\alpha_1|}^{1-{2\,\beta_1}}  }{
 \left( 2\,\alpha_1-\epsilon_1 \right) ^{2}}}\right){m}^{2\,\beta_1+1}
  ,\quad\,\,\,\,\,\,\,\quad\quad\quad\quad\,\quad\,\,\quad \quad\quad\quad\,\, m\ll\frac{1}{r_c},
\een
where we define the crossover scale $r_c=1/\epsilon_1$.
 And, the boundary conditions of the potential applied to the wavefunction (\ref{Bes12})
\be\label{Bound}
\psi_{m}\left(\frac{1}{\alpha_1}\right)=0,\quad z>0;\quad \psi_{m}\left(-\frac{1}{\alpha_1}\right)=0,\quad z<0
\ee allows us to determine how the parameter $\epsilon_1$ is related to the graviton masses, i.e.
\be\label{ENERGY}
m^2\sim\frac14\,\epsilon^2_{{1}}\left(\frac{J_{{\beta_{{1}}}} \left( {\frac {m}{\alpha_{{1}}}}   \right)  }{ J_{{\beta_{{1}}+1}} \left( {\frac {m}{\alpha_{{1}}}}   \right)  }\right)^2.
\ee
 For $|\alpha_1|\ll1$ (or in terms of the bulk cosmological constant $|\Lambda|\gg1$), in the asymptotic regime the solution of the Eq.~(\ref{ENERGY}) is approximately given by
\be\label{ENERGY.2}
\frac{2}{\epsilon_1\,m}={\rm{tg}}\left(\frac{m}{\alpha_1}-\frac{(2\,\beta_1+1)\pi}{4}\right)\quad\Rightarrow\quad m^2\simeq\frac{k^2\,\pi^2\,\alpha^2_1}{4},\quad k=1,2,3...\quad.
\ee In such regime, it is reasonable to
approximate the potential generated by discrete massive graviton states as a summation of Yukawa-like potentials, which makes the
total effective potential to have the form \cite{SDW,MM,SCH}
\be\label{03.fim2}
 V (r) = \psi^2_0(0)\frac{{\rm{e}}^{-m_0\,r}}{r}+\sum_{n}{\psi^2_n(0)}\frac{{\rm{e}}^{-m_0\,r}}{r},
\ee 
where the first term is contribution of zero mode and the second term corresponds to
the correction which is generated by the exchange of KK-modes.  However, the set of discrete states may be replaced by a continuous treatment for $|\alpha_1|\ll1$, such that $\sum_{n}\rightarrow\int$. In this case,  we have no zero mode. Then, the contribution for the potential cames from the massive modes (the second term of Eq.~\eqref{03.fim2}), which can be expressed by an integration in two distinct regions, as follows
\be\label{07.fim2}
\delta V(r)= \frac{M_5^{-3}}{r}\left({\frac {{2\,|\alpha_1|}^{1-{2\,\beta_1}}  }{
 \left( 2\,\alpha_1-\epsilon_1 \right) ^{2}}}\int_0^{\frac{1}{ r_c}} {   
   { {m}}  ^{2\,{{\beta_1}}+1}
\,{\rm e}^{-m\,r} }dm+{\frac {\pi }{4\,{\alpha^{2}_1}}}\int_{\frac{1}{ r_c}}^{\infty}\, \frac{{m}^{2}}{ \left( \epsilon_1^2+{ {4\,{m}^{2}}} \right)}\,
 {{\rm e}^{-m\,r}}\,dm\right)\,.
\ee
In the limit where the crossover scale is very small, i.e., $ r_c\to0$ ($\Delta\rightarrow-4+2\sqrt{2}$, $\frac{1}{\sqrt{|\Lambda|}}\rightarrow0$), the first integral in Eq.~(\ref{07.fim2}) is dominant, and we find the familiar behavior of Randall-Sundrum scenario \cite{RS1} 
  \ben\label{11.fim}
\delta V\left(r\right)\sim \frac{2\,|\alpha_{{1}}|\,{r_{{c}}}^{2} }{M^3_5\,\left( -1+2\,\alpha_{{1}}r_{{c}} \right) ^{
2}\,r^3} .
 \een  
{As we previously stated that $r_c\sim H^{-1}$, this regime is particularly achieved in the early Universe, where only very small distances can be probed.}

\begin{figure}[ht]
\includegraphics[{height=04cm,width=08cm,angle=00}]{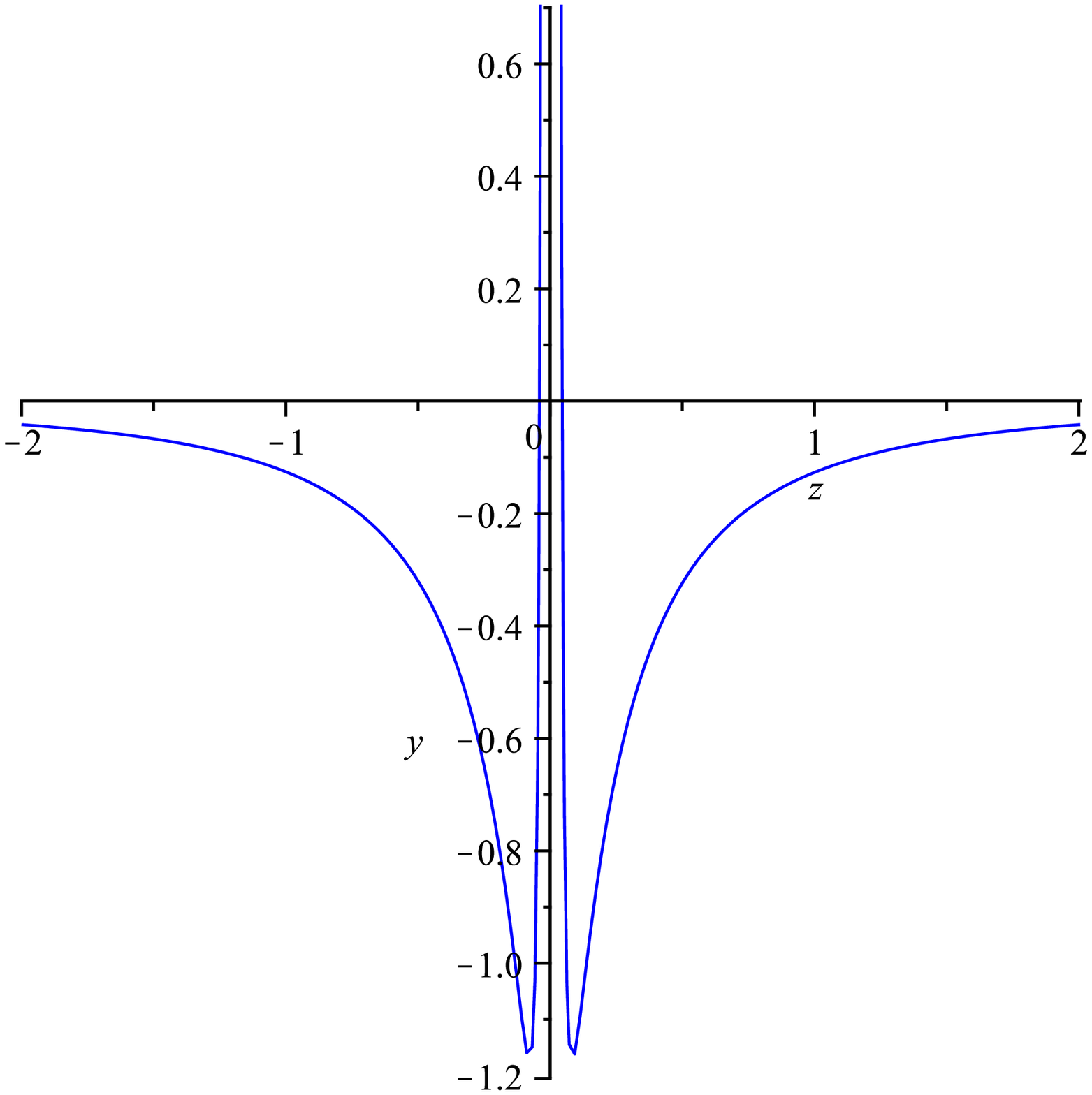}
\caption{The potential \eqref{dW20} as a function of $z$, for $\Delta=Q=1$.}\label{fig2}
\end{figure}

On the other hand, for the crossover scale being very large, i.e., $r_c\to\infty$ $(|\Delta|\gg Q)$, the second integral in Eq.~(\ref{07.fim2}) is dominant, and using the relation ${\rm E_1}(i\,x)=-{\rm ci}(x)+i\,{\rm si}(x)$, where ${\rm ci}(x)=-\int_x^{\infty}{dt\,\cos\,t/t}$ and ${\rm si}(x)=-\int_x^{\infty}{dt\,\sin\,t/t}$, we obtain the potential
\ben\label{CiSi.fim2}
\delta V \left( r \right) \sim{\frac {\pi }{4\,\alpha^2_1}
}
\left[{\frac {\,{{\rm e}^{-\frac{r}{2\,r_c}}}}{4\,{r}^{2}\,M_5^3\,\pi }}+ {\frac { \left( \sin \left( \frac{r}{2\,r_c}
 \right) {\rm ci} \left( \frac{r}{2\,r_c} \right) -\cos \left( 
\frac{r}{2\,r_c} \right) \,{\rm si} \left( \frac{r}{2\,r_c} \right) 
  \right)}{8\,r\,r_c\,\pi\,M_5^3 }}\right].
 \een
In order to analyze the behavior for large and small distance compared with the crossover scale $r_c$, we use the asymptotic forms: ${\rm ci}(x)\sim\gamma+\ln(x)$ and ${\rm si}(x)\sim x-{\pi}/{2}$ for $x\ll1$, $
{\rm ci}(x)\sim{{\rm sin}}(x)/{x}$ and ${\rm si}(x)\sim -{{\rm cos}}(x)/{x}$ for $ x\gg1$. 
For small distance, i.e., $r/r_c\ll1$  we obtain the following form 
\be\label{13.fim2}
\delta V \left( r \right) \sim 
\,{\frac {1}{32\,\alpha^2_1\, r_c\,{
M^{3}_{{5}}\,r}}
}\left(  \frac{r}{2\,r_c}\,\left( \gamma+\ln  \left( \frac{r}{2\,r_c}\right)  \right) -\frac{r}{2\,r_c}+\frac{\pi}{2}+{\cal O}(r^2)\right)\,,
\quad r\ll {r_c}.
\ee
where $\gamma\approx0.577$ is the Euler-Masceroni constant.
Notice that, at this limit  the potential has the correct $4D$ Newton's law with $1/r$
scaling.
Finally, for large distance, i.e., $r/r_c\gg1$, the potential in Eq.~(\ref{CiSi.fim2}) gives
\be\label{15.fim2}
\delta V \left( r \right) \sim \frac{1}{16}\left({\frac {1}{\alpha^2_1\,{M^{3}_{{5}}}\,{r}^{2}}}\right)\sim\frac{1}{r^2}
,\quad\quad r\gg {r_c}, 
\ee
that recovers the laws of $5D$ gravity \cite{RGS,DGP}.
\begin{itemize}
	\item ii) $\Delta>0$ case.\\
\end{itemize} 
In this case, for the potential in Eq.~(\ref{dW20}) 
the general solution of the Schroedinger-like equation \eqref{sch1} for massive modes is given by \cite{STE}
\be\label{01}
{\psi}_{m} \left( z \right) ={\rm{K}}_2(m)\,\sqrt {\frac{1}{k}+|z|} \left[J_{{\rm{n}}} \left( { {m \left( \frac{1}{k}+|z|
 \right) }} \right) -F_2(m)\,N_{{\rm{n}}} \left( { {m
 \left( \frac{1}{k}+|z| \right) }} \right)  \right] ,
\ee
where $J_{\rm{n}}$ and $N_{\rm{n}}$ are Bessel functions of the first and second kind, respectively, ${\rm{n}}=\frac14\,{\frac {{\Delta}^{2}}{ \left( \Delta+2 \right) ^{2}}}
$, $k={\frac { \left( \Delta+2 \right) Q}{\Delta}}$, and ${\rm{K}}_2$ and $F_2$ are the normalization factors --- See Fig.~\ref{fig2}. \\
The asymptotic behavior of $|{\psi}_{m}\left(0\right)|^2$ can be found from the
normalization factors
\be\label{Nor}
|{\rm{K}}_2(m)|^2=\frac{m}{2\,\left(1+F^{2}(m)\right)},
\ee
\be\label{F}
F_2\left(m\right)=\frac{\epsilon_2\,J_{{\rm{n}}} \left( \frac{m}{k} \right) +J_{{\rm{n}}+1}\left( \frac{m}{k} \right) m}{\epsilon_2\, N_{{\rm{n}}} \left( \frac {m}{k}
 \right) +N_{{\rm{n}}+1} \left( \frac{m}{k} \right) m},\quad \epsilon_{{2}}=k+2\,{\rm{n}}\,k+{\frac {2\,Q}{\Delta}}\,.
\ee
Another time, using the Lommel's formula, $J_{{\rm{n}}+1}(x)\,N_{{\rm{n}}}(x)-N_{{\rm{n}}+1}(x)\,J_{{\rm{n}}}(x)=2/\pi\,x$ and the asymptotic form of the Bessel functions, the probability density reduces to
\ben\label{Prob1}
\left| {\psi}_{m} \left( 0 \right)  \right|^2&\sim&\frac{2\,{m}^{2}}{{\pi }\, \left( \epsilon_2^2+{ {{m}^{2}}} \right)}
 ,\quad\,\,\,\quad\,\quad \quad\quad\quad\, \quad\quad\quad \quad\quad\quad\, \quad\,m\gg r_c\,,\\
\left| {\psi}_{m} \left( 0 \right)  \right|^2&\sim&  \frac{ 2 }{\left(\epsilon_2+1\right) ^{2}}\left( {\frac {m}{k}} \right) ^{2\,{\rm{n}}+1}
  ,\quad\,\,\,\,\,\,\,\quad\quad\quad\quad\,\quad\,\,\,\,\,\quad \quad\quad\quad\,\, m\ll r_c\,,
\een 
where we define the crossover scale $r_c=1/\epsilon_2$.

Here, to compute the correction to the four-dimensional Newtonian potential generated by the massive modes, we use the formula \cite{RS1}
\be\label{03}
V (r) = \frac{M^{-3}_{5}}{r}|\psi_0(0)|^2 + \delta V (r) ,
\ee 
where the first term is the contribution of the zero mode, and the second term corresponds to
the corrections generated by the exchange of massive Kaluza-Klein modes 
\be\label{intkk}
\delta V (r)= \frac{M_5^{-3}}{r}\int_0^\infty{dm|\psi_m(0)|^2\,e^{-mr}}\,.
\ee
As in the earlier case, there are no zero modes and we divide this integral into two regions limited by the crossover scale as
\be\label{07b}
\delta V(r)= \frac{M_5^{-3}}{r}\left(
\frac{ 2 }{\left(\epsilon_2+1\right) ^{2}}\int_0^{\frac{1}{ r_c}} {   
  \left( {\frac {m}{k}} \right) ^{2\,{\rm{n}}+1}
\,{\rm e}^{-m\,r} }dm+\int_{\frac{1}{ r_c}}^{\infty}\, \frac{2\,{m}^{2}}{\pi \, \left( \epsilon_2^2+{m^2}
\right)}\,
 {{\rm e}^{-m\,r}}\,dm\right)\,.
\ee
Thus, the Newton's law correction for massive modes at a distance $r$ is given as follows. 
In the limit where the crossover scale is very small, i.e., $ r_c\to0$ ($\Delta\ll Q$), the first integral in (\ref{07b}) is dominant and gives the familiar behavior of Randall-Sundrum scenario \cite{RS1}, for the critical values of the dilaton coupling $a^2\sim4\,\frac{(D-1)}{(D-2)^2}$ in this limit, so that
 \ben\label{11}
\delta V\left(r\right)\sim {\frac {2}{k\,(1+r_c)^2}
}\,\left(\frac{r^2_{{c}}}{{M^{3}_{{5}}}\,{r}^{3}}\right)\,.
 \een 
Now, for the crossover scale being very large, i.e., $r_c\to\infty$ $(\Delta\gg Q)$, the second integral in (\ref{07b}) is dominant and
we can get  $\delta V (r)$  approximately by \cite{RGS,FBL1}
\ben\label{CiSia}
\delta V \left( r \right) ={\frac {\,{{\rm e}^{-\frac{r}{r_c}}}}{{r}^{2}\,M_5^3\,\pi }}+ {\frac { \left[ \sin \left( \frac{r}{r_c}
 \right) {\rm ci} \left( \frac{r}{r_c} \right) -\cos \left( 
\frac{r}{r_c} \right) \,{\rm si} \left( \frac{r}{r_c} 
 \right)\right] }{2r\,r_c\,\pi\,M_5^3 }}\,.
 \een 
  
Once more, we use the asymptotic forms of ${\rm ci}(x)$ and ${\rm si}(x)$,  to examine the large and small distance behavior.
For small distance, i.e., $r/r_c\ll1$ we obtain the following form 
\be\label{13a}
\delta V \left( r \right) \sim \frac{1}{2\,r\,r_c{
M^{3}_{{5}}}\pi }
\,\left( \frac{r}{r_c}  \left( \gamma+\ln  \left( \frac{r}{r_c}\right)  \right) -\frac{r}{r_c}+\frac{\pi}{2}+{\cal O}(r^2)\right)\,, 
\quad r\ll{r_c}\,.
\ee
Interestingly, at short distance the computed potential has the correct $4D$ Newton's law with $1/r$
scaling. The next leading correction is given by the logarithmic repulsion term in (\ref{13a}).

On the other hand, for large distance, i.e., $r/r_c\gg1$, the potential in Eq.~(\ref{CiSia}) gives
 \be\label{15}
\delta V(r)\sim \,\frac {3}{2\,{r}^{2}\,{M^{3}_{{5}}}\,\pi }\,
,\quad\quad r\gg{r_c}\,,
\ee
which describes the laws of $5D$ gravity \cite{RGS,DGP}.

\begin{itemize}
	\item iii) $-1<\Delta<0$  case.
\end{itemize}
The following analysis is pretty similar to the previous one. In the present case, for the potential \eqref{dW20} written as
\be\label{molec}
U(z)=-\frac{{Q}^{2}}{{\Delta}^{2}}{\frac { \left( \Delta+1\right) }{\left(1-{\frac { \left( \Delta+2 \right)}{|\Delta|}}Q|z|\right)}}-\frac{2\,Q}{|\Delta|}\,\delta(z)\,,
\ee 
\begin{figure}[ht]
\includegraphics[{height=04cm,width=08cm,angle=00}]{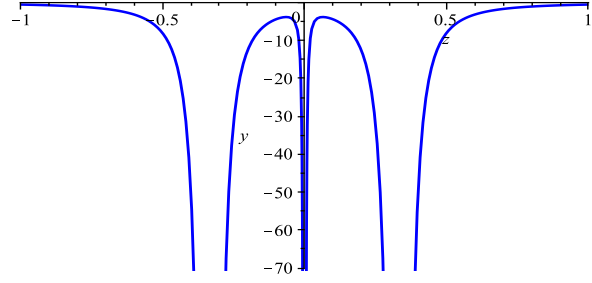}
\caption{The potential \eqref{dW20} as a function of $z$, for $\Delta=-1/2$, and $Q=1$.}\label{fig3}
\end{figure}
the Schroedinger-like equation gives the following solution
\be\label{Bes1}
{\psi}_{m} \left( z \right) ={\rm{K}}_3(m)\,\sqrt {\frac{1}{\alpha_2}+\left| z \right| }\,\left(J
_{{\beta_2}} \left( {\frac {m \left( 1+\alpha_2\, \left| z \right|  \right) 
}{\alpha_2}} \right) +F_3(m)\,N_{{
\beta_2}} \left( {\frac {m \left( 1+\alpha_2\, \left| z \right|  \right) }{
\alpha_2}} \right)\right)\,,
\ee 
for massive modes \cite{STE},
in terms of the Bessel functions of first and second kind, $J_{\beta_2}$ and $N_{\beta_2}$, respectively, where $\alpha_2=-{\frac {(2-\Delta)\,Q}{\Delta}}$, $\beta_2={\frac { \Delta }{2\,({\Delta}-2)}}$, and the normalization factors $K_3$ and $F_3$ --- See Fig.~\ref{fig3}.\\
The jump condition at $z = 0$ is
\be\label{JUMP.fim}
\left[{\frac {d}{dz}}{\psi}_{{m}} \left( z \right) \right
]_{{z=0}}=-\frac{2\,Q}{\Delta}\,{\psi}_{m}\left( 0 \right)\,.
\ee
The asymptotic behavior of $|{\psi}_{m}\left(0\right)|^2$ comes from the normalization factors given by
\be\label{NorFim}
|{\rm{K}}_3(m)|^{2}=\frac{m}{2\,|\alpha_2|\,\left(1+F_3^{2}(m)\right)},
\ee
\be\label{F.fim}
F_{{3}} \left( m \right) = \frac{\left( -2\,J_{{\beta_2+1}} \left( {\frac {m}{  \alpha_2}} \right) m+\epsilon_3 J_{{\beta_2}} \left( {\frac {m}{\alpha_2}  } \right)  \right) }{ \left( -2\,N_{{\beta_2+1}} \left( {\frac {m}{\alpha_2}  } \right) m+\epsilon_3 N_{{\beta_2}} \left( {\frac {m}{\alpha_2}}   \right)  \right) }\,,\quad\,\epsilon_3=2Q\,.
\ee 
Anew, using the Lommel's formula, $J_{{\rm{n}}+1}(x)\,N_{{\rm{n}}}(x)-N_{{\rm{n}}+1}(x)\,J_{{\rm{n}}}(x)=2/\pi\,x$ and the asymptotic form of the Bessel functions, the probability density become
\ben\label{Prob1.fim}
\left| {\psi}_{m} \left( 0 \right)  \right|^2&\sim&{\frac {\pi }{4\,{\alpha_2^{2}}}}\,\frac{{m}^{2}}{ \left( \epsilon_3^2+{ {4\,{m}^{2}}} \right)}
 ,\quad\,\,\,\quad\,\quad \quad\quad\quad\, \quad\quad\quad \quad\quad\quad \quad\quad\,m\gg\frac{1}{r_c}\,,\\
 \left| {\psi}_{m} \left( 0 \right)  \right| ^{2}&\sim&  \left({\frac {{2\,|\alpha_2|}^{1-{2\,\beta_2}}  }{
 \left( 2\,\alpha_2-\epsilon_3 \right) ^{2}}}\right){m}^{2\,\beta_2+1}\,,
 \quad\,\,\,\,\,\,\,\quad\quad\quad\quad\,\quad\,\,\quad \quad\quad\quad\,\, m\ll \frac{1}{r_c}\,,
\een 
where we define the crossover scale $r_c=1/\epsilon_3$.

Again, we have no zero mode, consequently, all  the contribution to the Newtonian potential comes from the massive modes given by Eq.~(\ref{intkk}), that we shall integrate in the two regions as
\be\label{07.fim}
\delta V(r)= \frac{M_5^{-3}}{r}\left({\frac {{2\,|\alpha_2|}^{1-{2\,\beta_2}}  }{
 \left( 2\,\alpha_2-\epsilon_3 \right) ^{2}}}\int_0^{\frac{1}{ r_c}} {   
   { {m}}  ^{2\,{{\beta_2}}+1}
\,{\rm e}^{-m\,r} }dm+{\frac {\pi }{4\,{\alpha}_2^{2}}}\int_{\frac{1}{ r_c}}^{\infty}\, \frac{{m}^{2}}{ \left( \epsilon_3^2+{ {4\,{m}^{2}}} \right)}\,
 {{\rm e}^{-m\,r}}\,dm\right),
\ee
When the crossover scale is very small, i.e., $ r_c\to0$ ($Q\rightarrow\infty$), the first integral in (\ref{07.fim}) is dominant and we naturally find the familiar behavior of the Randall-Sundrum scenario \cite{RS1} 
 \ben\label{11.fim}
\delta V\left(r\right)\sim \frac{2\,|\alpha_{{2}}|\,{r_{{c}}}^{2} }{M^3_5\,\left( -1+2\,\alpha_{{2}}r_{{c}} \right) ^{
2}\,r^3}.
 \een 
For the crossover scale being very large, i.e., $r_c\to\infty$ $(Q\to0)$ the second integral in (\ref{07.fim}) is dominant, and
using the relation ${\rm E_1}(i\,x)=-{\rm ci}(x)+i\,{\rm si}(x)$  we have \cite{RGS,FBL1}
\ben\label{CiSi.fim}
\delta V \left( r \right) \sim\frac14\,{\frac {\pi }{ \alpha_2^2}
}
\left({\frac {\,{{\rm e}^{-\frac{r}{2\,r_c}}}}{4\,{r}^{2}\,M_5^3\,\pi }}+ {\frac { \left[ \sin \left( \frac{r}{2\,r_c}
 \right) {\rm ci} \left( \frac{r}{2\,r_c} \right) -\cos \left( 
\frac{r}{2\,r_c} \right) \,{\rm si} \left( \frac{r}{2\,r_c} \right) 
 \right] }{8\,r\,\pi\,M_5^3\,r_c }}\right)\,.
 \een 
 \\
 Now, from the asymptotic form of ${\rm ci}(x)$ and ${\rm si}(x)$, we examine the large and small distance behavior comparing with the crossover scale. 
For small distance, i.e., $r/r_c\ll1$, we obtain
\be\label{13.fim}
\delta V \left( r \right) \sim 
\,{\frac {1}{32\, \alpha_2^2\,{
M^{3}_{{5}}\,r\,r_c}}
}\,\left( \frac{r}{2\,r_c} \left( \gamma+\ln  \left( \frac{r}{2\,r_c}\right)  \right) -\frac{r}{2\,r_c}+\frac{\pi}{2}+{\cal O}(r^2)\right)\,,
\quad r\ll{r_c}\,.
\ee
Notice that at short distances the potential has the correct $4D$ Newtonian $1/r$ scaling, with a subsequently correction by the logarithmic repulsion term in (\ref{13.fim}). \\
Finally, for large distance, i.e., $r/r_c\gg1$, the potential (\ref{CiSi.fim}) reduces to
 \be\label{15.fim}
\delta V(r)\sim \,{\frac {1}{16\,\alpha_2^2}}\,
\,\left(\frac{1}{{r}^{2}{M^{3}_{{5}}}}\right)
\sim\frac{1}{r^2}
,\quad\quad r\gg{r_c}, 
\ee
which is in accordance with the laws of $5D$ gravity \cite{RGS,DGP}. 

\begin{figure}[ht]
\includegraphics[{height=04cm,width=08cm,angle=00}]{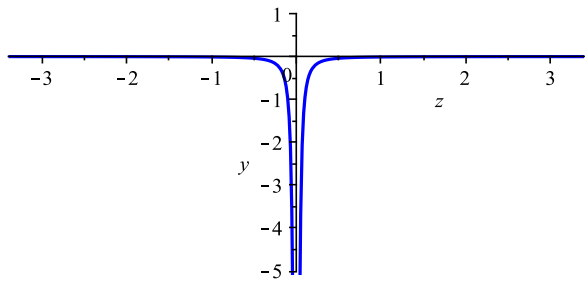}
\caption{The potential \eqref{dW20} as a function of $z$, for $\Delta=-1$, and $Q=1$.}\label{fig4}
\end{figure}

In the previous potentials all the scenarios with $r_c\to0$ implies large space curvature ($L\ll \ell_s$),  because $\Lambda$ becomes large and then the validity of supergravity approximation breaks down.
\begin{itemize}
	\item iv) $\Delta=-1$ case. 
\end{itemize}
In this case, the potential \eqref{dW20} reduces simply to $U(z)=-2\,Q\,\delta(z)$  --- See Fig.~\ref{fig4}.
The  probability density for
the massive modes is given in terms of
the scattering states governed by $|{\psi}_{m} \left( 0 \right) |^2$ that depends on the magnitude of the transmission $T$ or reflection $R$
coefficients. 
As usual, the jump condition at $z = 0$,
\be\label{JUMP}
\left[{\frac {d}{dz}}{\psi}_{m} \left( z \right) \right
]_{{z=0}}=-2\,Q\,{\psi}_{(m)} \left( 0 \right)
\ee
is obtained from the Schroedinger-like equation by using the properties of the delta function.

We now consider the general
wave functions for scattered states in the form
\ben\label{Wave}
\psi_{1{m}} \left( z \right)& =&{{\rm e}^{i\kappa\,z}}+R\,{{\rm e}^{-i
\kappa\,z}},\,\,\quad z<0\nonumber\\
\psi_{{2m}}\left( z \right)&=&T\,{{\rm e}^{i\kappa\,z}},\quad\quad\quad\quad\quad z>0
\een
where ${\kappa}$ is the wave number.
The simplified form for the probability density  in this case is
\ben\label{Prob1}
|T|^2=\left| {\psi}_{m} \left( 0 \right)  \right|^2={\frac {{m}^{2}}{\epsilon_4^2+{m}^{2}}}\,,\quad\,\epsilon_4=\sqrt{2\,\Lambda}\,.
\een 
Now we proceed as in the previous cases. Again, all the contribution to the Newton's law arises from the continuous massive modes on the brane, such that from \eqref{intkk} we have
\be\label{07}
\delta V(r)= \frac{M_5^{-3}}{r}
{\int_0^{\infty}} {   
  {\frac {{m}^{2}}{\epsilon_4^2+{m}^{2}}}
\,{\rm e}^{-m\,r} }dm\,.
\ee
Here, we define the crossover scale $r_c=1/\epsilon_4$, which imposes a bulk positive cosmological constant. 
For the crossover scale being very large, i.e., $r_c\to\infty$, we can get approximately the form of $\delta V (r)$ at a distance $r$, using the relation ${\rm E_1}(i\,x)=-{\rm Ci}(x)+i\,\left(-\pi/2+{\rm Si}(x)\right)$, where ${\rm Ci}(x)=-\int_x^{\infty}{dt\,\cos\,t/t}$ and ${\rm Si}(x)=\int_0^{\infty}{dt\,\sin\,t/t}$ \cite{DGP,FBL2}. The integral in (\ref{07}) turns
\ben\label{CiSik}
\delta V \left( r \right) ={\frac {1}{r\,r_c\,{M^{3}_{{5}}}}}\, \left[\sin
 \left( {\frac {r}{r_{{c}}}} \right) {\rm{Ci}} \left( {\frac {r}{r_{{c}
}}} \right) -\cos \left( {\frac {r}{r_{{c}}}} \right)\,\left(-\frac{\pi}{2}+ {\rm{ Si}} \left( 
{\frac {r}{r_{{c}}}} \right)\right)  \right] \,.
 \een
We now use the asymptotic forms: ${\rm Ci}(x)\sim\gamma+\ln(x)$ and ${\rm Si}(x)\sim x$ for $x\ll1$, $
{\rm Ci}(x)\sim{{\rm sin}}(x)/{x}$ and ${\rm Si}(x)\sim -{{\rm cos}}(x)/{x}$ for $ x\gg1$\,.\\
For small distance, i.e., $r/r_c\ll1$  we obtain the following form 
\be\label{13k}
\delta V \left( r \right) \sim \frac{1}{r\,r_c{
M^{3}_{{5}}} }
\,\left( \frac{r}{r_c} \left( \gamma+\ln \left( \frac{r}{r_c}\right)  \right) -\frac{r}{r_c}+\frac{\pi}{2}+{\cal O}(r^2)\right)\,,
\quad r\ll{r_c}\,.
\ee
As we have found, in the previous cases, at short distances the potential has the correct $4D$ Newtonian $1/r$
scaling.\\
Other wise, for large distance, i.e., $r/r_c\gg1$, the potential in Eq.~(\ref{CiSik}) gives
 \be\label{15k}
\delta V(r)\sim \,\frac {1}{{r}^{2}\,{M^{3}_{{5}}}\,\pi }
,\quad\quad r\gg{r_c}\,,
\ee
which naturally signals a $5D$ gravity behavior \cite{RGS,DGP}.

{The early potentials, for the cases (ii),  (iii), and  (iv),  require special attention due to the emergence of unbounded and/or negative Schroedinger-type
potential. The appearance of infinite continuous states, confined in the wells, can be controlled by an inversely proportional relationship with the parameter $Q$, which determines the width of each potential well. The calculation of the gravitational potential corrections concerning the non-normalizable unstable massive modes, i.e., the region where $-\infty<{m^2}<0$, can be effected by a change of parameter appropriate for each case. With the aim of verifying the characteristic of the respective gravity we redefine the parameter $k_b \rightarrow i \, k_b$ in the following analysis for large values of crossover scale $r_c$. Thus, these corrections  for the three cases take the general form}
\be\label{15.1}
\delta\,V^*\left( r \right) \sim C_b\,\int_{0}^{\infty}\, \frac{{m}^{2}}{r\,k_b^2\, \left( \epsilon_b^2+{ \frac{{{m}^{2}}}{k_b^2}} \right)}\,
 {{\rm e}^{-m\,r}}\,dm\rightarrow C_b\,\int_{0}^{\infty}\, \frac{{m}^{2}}{-r\,k_b^2\, \left( \epsilon_b^2-{ \frac{{{m}^{2}}}{k_b^2}} \right)}\,
 {{\rm e}^{-m\,r}}\,dm\,,\quad b=2,3,4\,,
\ee 
{where $k_2=k$, $k_3=1/2$, $k_4=1$, and  $\epsilon_b$ were defined in  \eqref{F},  \eqref{F.fim}, and \eqref{Prob1}.}

{Let us analyze the behavior for  large and small distance  with respect to the scale $r_c=1/\epsilon_b$.  When $r_c\rightarrow 0$, there is no additional contribution for potential  coming from  Eq.~(\ref{15.1}), invading the domain investigated in (ii), (iii), and (iv) cases. For $r_c\rightarrow \infty$, using the relation ${\rm E}_1(x)\sim{\rm Shi}(x)-{\rm Chi}(x)$, ${\rm Chi}(x)=\gamma+\ln(x)+\int_0^{x}{dt\,(\cosh\,t-1)/t}$ and ${\rm Shi}(x)=\int_0^{x}dt\,{\sinh\,t/t}$, the Eq.~(\ref{15.1}) provides us the dominant term for the correction of the gravitational potential as}
\ben\label{CiSia-1}
\delta\,V^*\left( r \right) \sim C_{{b}} \left[ \frac{1}{{r}\,{r_{{c}}}}\,\left({\rm Chi} \left( {\frac {r}{r_{{c}}}} \right) \sinh
 \left( {\frac {r}{r_{{c}}}} \right) -\cosh
 \left( {\frac {r}{r_{{c}}}} \right) {\rm Shi} \left( {\frac {r}{r_{{c
}}}} \right) \right) \right]\,.
 \een 
{Now, we use the asymptotic forms: ${\rm Chi}(x)\sim\gamma+\ln(x)$, ${\rm Shi}(x) \sim x$, $\sinh(x)\sim x+\frac{x^3}{6}$ and $\cosh(x)\sim 1+\frac{x^2}{2}$ for $x\ll1$; ${\rm Chi}(x)\sim{{\rm sinh}}(x)/x$ and ${\rm Shi}(x)\sim-1/2\,\pi+{{\rm cosh}}(x)/x$ for $ x\gg1$. 
For small distance, i.e., $r/r_c\ll1$ we obtain the corrections given by}
\be\label{13b}
\delta V^*\left( r \right) \sim -\frac{C_b}{r^2_c}
\,\left(  \gamma+\ln  \left( \frac{r}{r_c}\right)   -1+{\cal O}(r^2)\right)\,;
\ee
{where  $C_2=1/({2\,\pi\,M^3_5})$, $C_3=1/({32\,\alpha_2^2\,M^3_5})$, and  $C_4=1/{M^3_5}$,
for the cases (ii),  (iii), and (iv), respectively.}

{ Finally, the resulting potentials are given by the sum of the correction \eqref{13b} with \eqref{13a},  \eqref{13.fim}, and \eqref{13k},  for the cases (ii), (iii), and (iv), respectively.
 This analysis always involves the relationship $r/r_c\ll1$, concerning the three cases studied and, interestingly, at short distance the computed potential has the correct \textit{attractive} $4D$ \textit{Newton's law} with $1/r$ scaling.   
On the other hand, for large distance, i.e., $r/r_c\gg1$, the potential in Eq.~(\ref{CiSia-1}) gives
\be\label{15}
\delta V^*(r)\sim -\frac{C_b}{r^2}\,\left[\frac12\, \left( -2+\cosh \left( {\frac {r}{r_{{c}}}} \right) \pi 
 \right) \right]
\sim-\frac{C_b}{4\,r^2}\,\pi\,{\rm e}^{{\frac {r}{r_{{c}}}}}
,\quad\quad b=1,2,3; \quad r\gg{r_c},
\ee
which implies a scenario of dominant repulsive gravity. The exponential increasing of the potential with $r$ also signals a {\it gravitational confining phase} where any particle has infinite energy. As a consequence of such a confinement, no isolated particles can be found such that this
phase indeed comprehends a vacuum state in a curved space which is able to produce an inflationary phase of the Universe \cite{linde}.

This is in accord with the fact that plugging $m^2\to -m^2$ into Schroedinger-like equation (\ref{sch1}) one finds formally the same equation as long as $z\to -iz$, i.e., the extra-dimension becomes a time-like one, and the potential in (\ref{dW20}) goes like $U(z)\to -U(z)$. This, of course, implies $Q\to iQ$  (in the bulk, i.e., $z\neq0$) that from Eq.~(\ref{dW6b}) flips the sign of the bulk cosmological constant $\Lambda$. This change of sign leads to a positive cosmological constant and then to a five-dimensional de Sitter space ($dS_5$) which is indeed the aforementioned vacuum developing an inflationary phase of the Universe.  This is closely related to earlier studies on time-like extra-dimensions \cite{chaichian, Matsuda:2000nk}.

Now, before ending this section, some brief comments about the limit of very small crossover scale are in order.  When $r_c\to0$, the additional contribution to the Newtonian potential is dominated by integrals of the type $\frac{1}{r}\int_0^\infty{m^{2n+1} e^{-mr}dm}$. They impose  some restriction on the non-normalizable unstable modes $-\infty<m^2<0$. This can be easily seen by changing the variables $m\to -im$ and $r\to ir$ to see that the integral get one complex factor $i^{2n+1}$. Recalling  that $|\psi|^2\sim m^{2n+1} $ is always positive, the power of this factor can only assume the values:
$2n+1=4,8,...$. This implies attractive gravity depending on each parameter $n$ of the respective cases (ii) and (iii) considered above.
}

\section{Conclusions}\label{SecIV}

In this paper, by considering massive graviton modes coming from a dilatonic domain wall solution of five-dimensional supergravity, we have shown that the gravitational potential corresponds to the usual Newton potential which scales with $~1/r$ at short distance and has a five-dimensional behavior scaling with $~1/r^2$ at large distance compared with the crossover scale $r_c$.{We have explored distinct cases which depend on the parameter $\Delta$. In our present analysis,  we have chosen to work out with $\Delta$ in four different ranges. We have shown that, in all of these cases is recovered the four dimensional attractive gravity in the regime $r\ll r_c$. The first case gives the expected result since the localized gravity is attractive everywhere, including the five-dimensional regime at  $r\gg r_c$. However, due to non-normalizable unstable massive graviton modes present on the spectrum, the following three cases assume a new behavior at large distance, i.e., at $r\gg r_c$. At this regime they exhibit a repulsive gravity with exponentially increasing potential which signals a gravitational confining phase that is able to produce an inflationary phase of the Universe \cite{linde}}. This has also been addressed in earlier studies on time-like extra-dimensions \cite{chaichian, Matsuda:2000nk}. This study showed that from a five-dimensional supergravity theory with a scalar field describing the dilaton, the emergence of 
four-dimensional gravity on a $3$-brane is possible even if the brane is embedded in  an asymptotic five-dimensional flat space, below a crossover scale and the manifestation of extra dimensions does not necessarily occur only at short distances as commonly expected. Localization of gravity in other dilatonic domain walls configurations has also been addressed in \cite{FBL2,DI6,DI7}. 

\acknowledgments
The authors would like to thank CAPES,
CNPq, CAPES/PROCAD/PNPD and PRONEX/CNPq/FAPESQ for partial support.



\end{document}